\documentclass[journal=jpcafh,manuscript=article]{achemso}
\usepackage{longtable}
\usepackage{multirow}
\usepackage{amsmath}
\usepackage{subfigure}
\usepackage[usenames,dvipsnames]{color}
\usepackage{soul}
\usepackage{bm}

\usepackage{xr}
\usepackage{amssymb}

\usepackage{multicol} \usepackage{wrapfig} \usepackage{enumitem}
\usepackage{bm} \usepackage{outlines} %for itemize with subitems
\SectionNumbersOn

\author{Silvan K\"aser} \affiliation[University of Basel]{Department of
  Chemistry, University of Basel, Klingelbergstrasse 80 , CH-4056
  Basel, Switzerland.}
\author{Markus Meuwly} \affiliation[University of Basel]{Department of
  Chemistry, University of Basel, Klingelbergstrasse 80 , CH-4056
  Basel, Switzerland.}  \email{m.meuwly@unibas.ch}

\title{Transfer Learned Potential Energy Surfaces: Accurate Anharmonic
  Vibrational Dynamics and Dissociation Energies for the Formic Acid
  Monomer and Dimer}

\begin{document}

\date{\today}\\

\begin{abstract}
The vibrational dynamics of formic acid monomer (FAM) and dimer (FAD)
is investigated from machine-learned potential energy surfaces at the
MP2 (PES$_{\rm MP2}$) and transfer-learned (PES$_{\rm TL}$) to the
CCSD(T) levels of theory. The normal modes and anharmonic frequencies
of all modes below 2000 cm$^{-1}$ agree favourably with experiment
whereas the OH-stretch mode is challenging for FAM and FAD from normal
mode analyses and finite-temperature MD simulations. VPT2 calculations
on PES$_{\rm TL}$ for FAM reproduce the experimental OH frequency to
within 22 cm$^{-1}$. For FAD the VPT2 calculations find the
high-frequency OH stretch at 3011 cm$^{-1}$, compared with an
experimentally reported, broad ($\sim 100$ cm$^{-1}$) absorption band
with center frequency estimated at $\sim 3050$ cm$^{-1}$. In agreement
with earlier reports, MD simulations at higher temperature shift the
position of the OH-stretch in FAM to the red, consistent with improved
sampling of the anharmonic regions of the PES. However, for FAD the
OH-stretch shifts to the blue and for temperatures higher than 1000 K
the dimer partly or fully dissociates using PES$_{\rm TL}$. Including
zero-point energy corrections from diffusion Monte Carlo simulations
for FAM and FAD and corrections due to basis set superposition and
completeness errors yield a dissociation energy of $D_0 = -14.23 \pm
0.08$ kcal/mol compared with an experimentally determined value of
$-14.22 \pm 0.12$~kcal/mol.
\end{abstract}

\section{Introduction}
Vibrational spectroscopy is a powerful means to relate structure and
dynamics of molecules in the gas phase and in solution. Combined with
state-of-the art simulation techniques, molecular-level information
such as assignments of local reporters, or the thermodynamics in
protein-ligand complexes can be obtained. One of the pertinent
questions is a) whether or not classical molecular dynamics (MD)
simulations can be used for this and b) what level of theory and
accuracy in representing the energies from electronic structure
calculations is required for a meaningful contribution of simulations
to assigning and interpreting experimentally determined
spectra. Electronic structure calculations and their representations
as full-dimensional potential energy surfaces (PESs) have progressed
to a degree that now allows direct comparison with
experiments.\cite{MM.sd:2017,MM.hyperson:2020,mm.ht:2020}\\

\noindent
Generating full-dimensional, reactive PESs even for small molecules is
a challenging task.\cite{meuwly2021machine,unke2021machine} This often
requires datasets consisting of tens of thousands of \textit{ab
  initio} calculations to adequately describe configurational space of
the system of interest. While calculations at the density functional
theory (DFT) or M{\o}ller-Plesset perturbation (MP2) levels of theory
are cost-efficient, reaction barriers are less
accurate\cite{qu2021breaking}. On the other hand, the ``gold
standard'' coupled cluster with perturbative triples (CCSD(T))
approach scales as $N^7$ (with $N$ being the number of basis
functions)\cite{friesner2005ab} which becomes quickly computationally
prohibitive for full-dimensional PESs even for moderately sized
molecules ($N_{\rm atoms} \sim 10$).  Recent applications of novel
machine learning (ML) approaches combined with transfer learning
(TL)\cite{pan2009survey,taylor2009transfer,smith2018outsmarting} and
related $\Delta$-ML\cite{DeltaPaper2015} in physical/computational
chemistry\cite{smith2018outsmarting,mm.ht:2020,mm.vibratingh2co:2020,nandi2021delta,mm.anharmonic:2021,qu2021breaking}
were shown to be data and cost-effective alternatives which will also
be explored in the present work.\\

\noindent
Formic acid (HCOOH), the simplest carboxylic acid, is an important
intermediate in chemical synthesis and relevant to atmospheric
chemistry\cite{khare1999atmospheric}. It is also a promising fuel and
H$_2$ carrier\cite{van2019fuelling,wang2020molecularly} which can be
produced by electrocatalytic CO$_2$ reduction reactions and, thus, may
contribute to decreasing atmospheric CO$_2$
levels.\cite{wang2020molecularly,xia2019continuous} Formic acid
monomer (FAM) and its dimer (formic acid dimer, FAD) have been the
subject of several
experimental\cite{ito2000jet,freytes2002overtone,georges2004jet,zielke:2007,xue2009probing,suhm:2012,MM.fad:2016,suhm:2020}
and
theoretical\cite{kalescky2013local,ivanov2015quantum,miliordos2015validity,tew2016ab,qu2016ab,MM.fad:2016,richardson2017full,qu2018high,qu2018quantum,qu2018ir}
studies. In the vapor phase, formic acid exists as hydrogen bonded
dimers\cite{reutemann2011formic} making it a prototype for complexes
with hydrogen bonds such as enzymes or DNA base
pairs\cite{balabin2009polar}. The experimental IR spectrum has been
reported for both, FAM and
FAD\cite{millikan1957fam,luiz1997gas,nejad2020increasing,millikan1958fad,MM.fad:2016,georges2004jet,ito2000jet,ito2002jet,meyer2020shifting}.\\

\noindent
For FAM a recent study of both, cis- and trans-FAM, analyses a global
PES constructed using regression techniques with energies from
CCSD(T)(F12*)/cc-pVTZ-F12 calculations as the
reference\cite{tew2016ab}. Vibrational eigenstates for both conformers
were calculated using vibrational configuration interaction (VCI) and
the fundamentals for trans-FAM were found to be in close agreement
with experiment (${\rm RMSD}=3$~cm$^{-1}$). Subsequently, a second
full-dimensional CCSD(T)-F12a/aug-cc-pVTZ based PES for the two FAM
conformers was used for multi-configuration time-dependent Hartree
(MCTDH) vibrational calculations\cite{richter2018vibrational}. The
trans fundamental transitions were found to agree to within
5~cm$^{-1}$ with experiment. The two
PESs\cite{tew2016ab,richter2018vibrational} are further analyzed in
detail in Ref.~\cite{nejad2021raman} and are used for comparison with
experiments for FAM and deuterated isotopologues of both
isomers. Vibrational eigenstates were obtained from high order
canonical Van Vleck perturbation theory (CVPT). Such high level
anharmonic treatments of vibrations on high quality PESs are crucial
for understanding and potentially (re)assigning experimental spectra
as was found for
trans-FAM\cite{tew2016ab,richter2018vibrational,nejad2021raman}:
assignment of the O-H in plane bend $\nu_5$ which is in resonance with
the O-H torsional overtone $2\nu_9$ has been ambiguous for a long
time. Based on the high-level calculations the fundamental was
assigned to 1306~cm$^{-1}$ and the overtone to a spectroscopic feature
at 1220~cm$^{-1}$. This assignment was subsequently supported by Raman
jet experiment\cite{nejad2020increasing}. Interestingly, this
reassignment was recently supported from second order vibrational
perturbation theory (VPT2) calculations using a neural network-(NN)
based PES.\cite{mm.anharmonic:2021}\\

\noindent
For FAD a considerable amount of computational work has been done on
its dynamics\cite{kim1996direct,miura1998ab}, infrared (IR)
spectroscopy\cite{MM.fad:2016} and calculations of tunneling
splittings.\cite{richardson2017full,qu2016ab} A recent PES based on
13475 CCSD(T)-F12a/haTZ energies was fit to permutationally invariant
polynomials\cite{qu2016ab} which reported a barrier for double proton
transfer of 2853~cm$^{-1}$ (8.16~kcal/mol). This was used for
computing the zero-point energy (ZPE) based on diffusion Monte Carlo
(DMC) calculations, for vibrational self consistent field (VSCF)/VCI
calculations of fundamentals and to determine ground-state tunneling
splittings in a reduced-dimensional approach. The tunneling splitting
obtained was 0.037~cm$^{-1}$ which is larger than the experimentally
reported value\cite{ortlieb2007proton,goroya2014high} of
0.016~cm$^{-1}$. Subsequently,\cite{richardson2017full} the same PES
was used for computing tunneling splittings without reducing the
dimensionality of the problem using the ring-polymer instanton
approach.\cite{richardson2011ring} The value of 0.014~cm$^{-1}$ was
considerably closer to experiment and a more recent measurement based
on IR techniques reported a tunneling splitting of
0.011~cm$^{-1}$.\cite{duan:2017}\\

\noindent
More recently the CCSD(T)-F12a PES was extended with a new dipole
moment surface (DMS) based on the MP2/haTZ level of theory and used
for the analysis of the FAD IR spectrum.\cite{qu2018high} Both, VSCF
and VCI calculations were performed and compared with results from
classical and ''semi-classically`` prepared quasiclassical MD
simulations and experiment.\cite{qu2018quantum} Classical MD
corresponds to \textit{NVE} simulations run at 300~K whereas the
harmonic ZPE is added in the ``semi-classically'' prepared approach.
For both approaches, the X-H stretch frequencies remain at higher
frequencies compared to experiment. This effect is ascribed to the
inability of MD simulations to correctly sample anharmonicities. The
high-quality PES and DMS\cite{qu2018high} were recently employed to
assess the fingerprint region of FAD using variational vibrational
computations and curvilinear kinetic energy operator
representation.\cite{santa2021fingerprint}\\

\noindent
Experimentally, the dissociation energy of FAD into two FAMs has been
determined from spectroscopy and statistical thermodynamics to be $D_0
= 59.5(5)$ kJ/mol ($\sim 14.22$ kcal/mol).\cite{suhm:2012} For the
double proton transfer barrier, the most recent value from microwave
spectroscopy (measured tunneling splitting of 331.2 MHz) is 2559
cm$^{-1}$ (30.6 kJ/mol or 7.3 kcal/mol) from analysis of a 3D
model.\cite{caminati:2019} This is consistent with a best value of 7.2
kcal/mol from a morphed MP2 surface that was determined from atomistic
simulations and compared with IR experiments of the proton transfer
band.\cite{MM.fad:2016} Earlier experiments reported a larger
splitting of 0.0158 cm$^{-1}$ (corresponding to 473.7 MHz) which
yields a somewhat higher barrier for double proton
transfer.\cite{ortlieb2007proton} Yet more recent work in the infrared
found 0.011367 cm$^{-1}$ (340.8 MHz) which is closer to the most
recent microwave data.\cite{duan:2017}\\

\noindent
In order to assess the expected magnitude of error cancellation due to
shortcomings in the electronic structure method used and that fact
that MD-based spectroscopy only samples the bottom of the potential
well at ambient conditions, extensive higher-level calculations at the
MP2 and CCSD(T) levels of theory were carried out together with the
aug-cc-pVTZ basis set. This reference data was then represented as a
neural network (NN) based on the PhysNet
architecture.\cite{MM.physnet:2019} Additionally, TL
\cite{smith2018outsmarting,pan2009survey,taylor2009transfer} to the
the CCSD(T)/aug-cc-pVTZ level of theory was used to further improve
the quality of the PES. Using these PESs for formic acid monomer and
dimer, the harmonic and anharmonic vibrations, and the IR spectrum
from finite-temperature MD simulations are determined. From this the
complexation-induced red shifts can be determined and compared with
experiment. Furthermore, DMC simulations are run to obtain an estimate
for the ZPE of both molecules. The DMC calculations are a meaningful
probe for the robustness of the PES and the resulting ZPEs are used
for the determination of the binding energy of the dimer.\\

\noindent
First, the methods are presented and discussed. Next, the accuracy of
the NN-based PESs is assessed and the vibrational spectra for FAM and
FAD are determined and compared with experiment. Then, the results
from DMC simulations are presented and, finally, the results are
discussed and conclusions are drawn.\\

\section{Computational Details}
\label{sec:Methods}
\subsection{Electronic Structure Calculations and PhysNet Representation}
The PESs for FAM and FAD are represented by a NN of the PhysNet
architecture\cite{MM.physnet:2019}, which has been used for different
chemical systems recently
\cite{MM.diels:2019,brickel2019reactive,mm.atmos:2020,MM.mgo:2020,mm.ht:2020,mm.vibratingh2co:2020}.
PhysNet is a high-dimensional NN\cite{behler2007generalized} of the
``message-passing'' type\cite{gilmer2017neural}. It constructs a
feature vector for each atom in a molecule describing the atom's local
chemical environment. These feature vectors (descriptors) are
iteratively refined (learned) and used to predict atomic contributions
to the total energy $E$ and partial charges $q_i$.  The forces
$\bm{F}_i$ needed to run MD simulations are obtained from reverse-mode
automatic differentiation\cite{baydin2017automatic} and the molecular
dipole moment is calculated from the partial charges following
$\bm{\mu} = \sum_i q_i \bm{r}_i$. Details of the PhysNet approach are
described elsewhere\cite{MM.physnet:2019}.  The parameters of PhysNet
are fitted to \textit{ab initio} energies, forces and dipole moments
calculated at the
MP2\cite{moller1934note}/aug-cc-pVTZ\cite{kendall1992electron} level
of theory using Molpro\cite{MOLPRO}.\\

\noindent
A dataset of reference structures is required for training PhysNet. An
ensemble of 20000 FAM (5000) and FAD (15000) geometries are generated
from Langevin dynamics at 2000~K using the atomic simulation
environment (ASE)\cite{larsen2017atomic} at the PM7 level of
theory\cite{stewar2016mopac, stewart2007optimization}. For FAD, the
transition state (TS) region is sampled by harmonically biasing the
geometry toward the TS structure, similar to the umbrella sampling
approach \cite{torrie1977nonphysical}. This dataset is extended with
geometries of fragments of the FAM molecule (H$_2$, CH$_4$, H$_2$O,
CO, H$_3$COH, H$_2$CO) following the amons
approach\cite{huang2020quantum}. For the amons, 1000 geometries each
are generated using Langevin dynamics at 1000~K. Based on initial
PhysNet representations of the PES, the dataset was extended with one
round of adaptive
sampling\cite{behler2016perspective,behler2015constructing}.  The
final MP2 dataset contains 26000 reference structures and was split
according to 20800/2600/2600 for
training/validation/testing. Henceforth, this PES is referred to as
PES$_{\rm MP2}$.\\

\noindent
To further improve the quality of the PES,
TL\cite{smith2018outsmarting,pan2009survey,taylor2009transfer} from
the MP2 to the CCSD(T) level of theory was performed. The data set for
TL contained 866 geometries: 425 for FAM and 441 for FAD. For FAM,
they were generated from normal mode sampling\cite{smith2017ani} at
different temperatures (between 10 and 2000~K) and geometries for FAD
were those along the minimum energy path (MEP), along particular
normal modes and geometries obtained from normal mode
sampling. \textit{Ab initio} energies, forces and dipole moments for
the 866 geometries were determined at the
CCSD(T)/aug-cc-pVTZ\cite{pople1987quadratic,purvis1982full,kendall1992electron}
level of theory using MOLPRO\cite{MOLPRO}. Then, the PhysNet model was
retrained on the TL-data set by initializing the NN with the
parameters from the trained NN at the MP2 level as a good initial
guess. The data set was split randomly according to 85/10/5~\% into
training/validation/test set and the learning rate was reduced to
$10^{-4}$ compared with $10^{-3}$ when learning a model from scratch.
The final TL PES is referred to as PES$_{\rm TL}$.\\

\subsection{Vibrational Calculations}
Harmonic and anharmonic frequencies can be determined using PES$_{\rm
  MP2}$ and PES$_{\rm TL}$. First, harmonic frequencies are computed
from diagonalization of the mass-weighted Hessian matrix. Next,
anharmonic frequencies are obtained (i) from finite-temperature IR
spectra calculated from the dipole-dipole autocorrelation
function\cite{thomas2013computing,schmitz2004vibrational,schmitz2004vibrational2}
of MD simulations and (ii) from VPT2 as implemented in the Gaussian
software\cite{barone2005anharmonic} using PES$_{\rm MP2}$ and
PES$_{\rm TL}$ as external potentials.\\

\subsection{Diffusion Monte Carlo}
DMC calculations are used to determine ZPEs (i.e. the quantum ground
state energy) for FAM and FAD, and to examine PES$_{\rm TL}$ for
holes. Here, the unbiased DMC algorithm was
used\cite{anderson1975random,kosztin1996introduction,qu2021breaking}.
A set of thousands to tens of thousands of random walkers is
initialized and subsequently the atoms in each walker are displaced
randomly at every timestep. The ensemble of walkers represents the
nuclear wavefunction of the molecule. Based on a walker's potential
energy $E_i$ with respect to a reference energy, $E_r$, they remain
alive and can give birth to new walkers, or can be killed. The walkers
die and replicate according to the following
probabilities\cite{qu2021breaking}:
\begin{align}
    P_{\rm death} = 1 - e^{-(E_i -E_r)\Delta\tau} \quad (E_i > E_r)\\
    P_{\rm birth} = e^{-(E_i -E_r)\Delta\tau} - 1 \quad (E_i < E_r).
\end{align}
Here, $\Delta\tau$ is the step size in imaginary time. Following each
time step, the dead walkers are removed and $E_r$ is updated according
to
\begin{align}
    E_r(\tau) = \left<V(\tau)\right> - \alpha \frac{N(\tau) -
      N(0)}{N(0)}
\label{eq:update_dmc}
\end{align}
In Eq.~\ref{eq:update_dmc} $\left<V(\tau)\right>$ is the averaged
potential energy over the alive walkers, $\alpha$ is the magnitude of
fluctuations in the number of walkers and is a parameter, and $N(0)$
and $N(\tau)$ are the number of alive walkers at times 0 and $\tau$,
respectively. The ZPE is then approximated as the average of $E_r$
over all imaginary times.\\

\noindent
Following Ref.~\citenum{qu2021breaking}, the DMC calculations were
performed in Cartesian coordinates and in full dimensionality. For
both, FAM and FAD, ten independent DMC simulations were performed. To
obtain an estimate of the ZPEs 30000 walkers were advanced for 50000
time steps following an equilibration period of 5000 time steps. A
step size of $\Delta\tau = 5.0$~au was used. For converging the DMC
simulations outlined above and to obtain statistically meaningful
results more than $3\cdot10^{10} (= 30000 \times 55000 \times 10
\times 2)$ energy evaluations were required in total. The
computational efficiency of a PES is crucial for DMC simulations which
requires a large number of energy evaluations. The Tensorflow
library\cite{abadi2016tensorflow}, in which PhysNet is implemented,
allows training and evaluation of the NN on graphics processing units
(GPUs). The use of GPUs for the evaluation of PhysNet speeds up the
calculations as they can be performed in a highly parallel
manner\cite{dirisio2021gpu}. Here, all DMC calculations were performed
on a GeForce RTX 2080Ti GPU with 12~Gb of RAM.  Each DMC simulation
for FAM (FAD) takes 8.7 (23.7)~h on average.\\

\section{Results}
\subsection{Quality of the PhysNet PES$_{\rm MP2}$ and PES$_{\rm TL}$}
\label{sec:quality}
\begin{figure}[h!]
\centering
\includegraphics[width=0.9\textwidth]{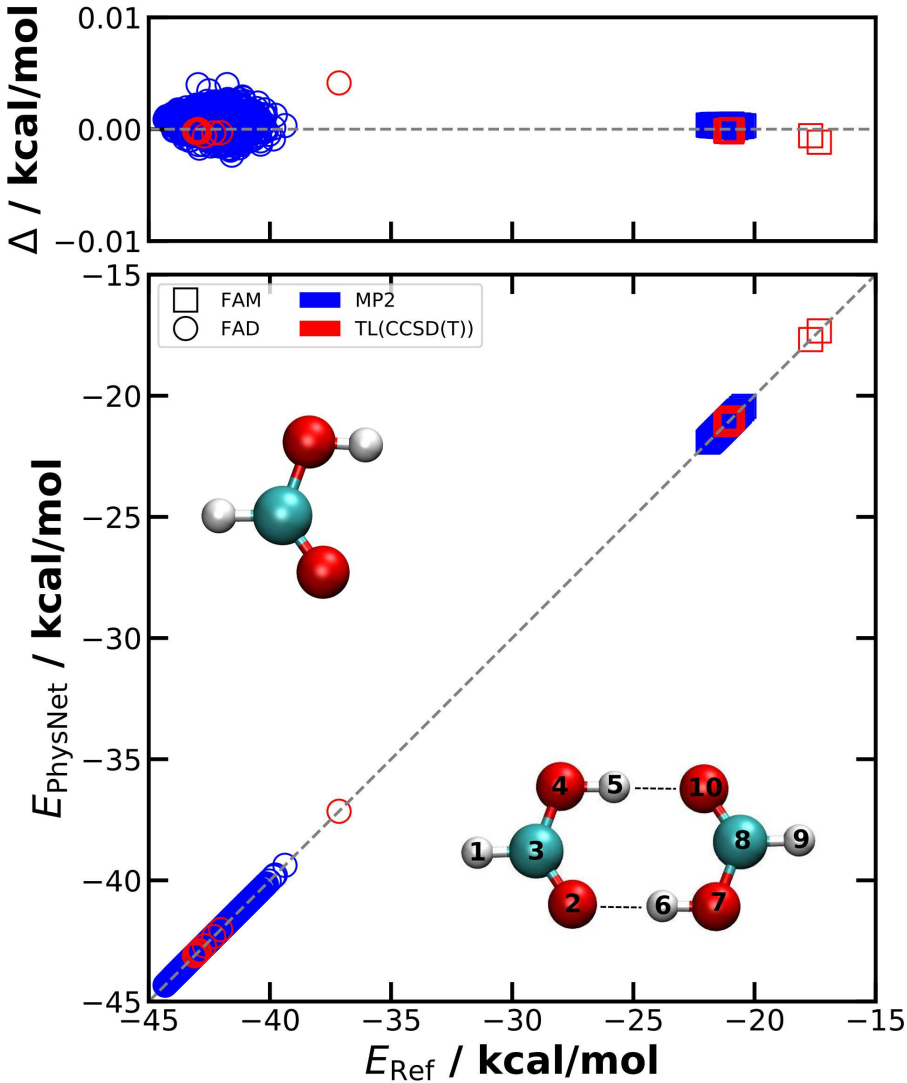}
\caption{The accuracy of the PES$_{\rm MP2}$ and PES$_{\rm TL}$
  energies is shown with respect to the appropriate reference
  \textit{ab initio} values taken from the separate test sets. For
  PES$_{\rm MP2}$, only the performance of the superior model is shown
  and only FAM and FAD geometries of the test set are considered. The
  deviations $\Delta = E_{\rm Ref} - E_{\rm PhysNet}$ for geometries
  in the test sets remain well below 0.01~kcal/mol. The structures of
  FAM and FAD are shown together with labels and the hydrogen bonds
  indicated by dotted lines.}
\label{fig:ooserrors}
\end{figure}

First, the quality of the NN-learned energy functions is
considered. To assess the reproducibility of the PhysNet models, two
independent models are trained for the MP2 reference data. Both are
evaluated on a test set containing 2600 randomly chosen structures of
the dataset which were not used during training. The model yielding
lower mean absolute and root mean squared errors (MAE and RMSE) on the
test set is then used for the simulations and for TL. The out of
sample performances of both PES$_{\rm MP2}$ and PES$_{\rm TL}$ are
summarized in Table~\ref{tab:oos_errors} and
Figure~\ref{fig:ooserrors}.\\

\begin{table}[h]
\begin{tabular}{l|c|c|c}
& \textbf{PES1$_{\rm MP2}$} & \textbf{PES2$_{\rm MP2}$} & \textbf{PES$_{\rm TL}$} \\\hline
\textbf{MAE($E$)}       & 0.012       & 0.014 & 0.007        \\
\textbf{RMSE($E$)}       & 0.019       & 0.028 & 0.016      \\
\textbf{MAE($F$)}       & 0.026       &  0.029 & 0.076      \\
\textbf{RMSE($F$)}       & 0.195       &  0.411 & 0.580      \\
\textbf{MAE($\mu$)}       & 0.001       & 0.001 & 0.002        \\
\textbf{RMSE($\mu$)}       & 0.002       & 0.002 & 0.008      \\\hline
\textbf{1-$R^2$($E$)} & 3.7E-9 & 8.5E-9 & 4.1E-9
\end{tabular}
\caption{Out of sample errors for two MP2 and the TL models. Both
  PESs$_{\rm MP2}$ are evaluated on a test set containing 2600
  randomly chosen structures of the dataset which were not used during
  training. PES1$_{\rm MP2}$ is chosen for further analysis due to
  better overall performance. PES$_{\rm TL}$ is tested on a separate
  test set containing 44 geometries. The energy, force and dipole
  moment errors are given in kcal/mol, kcal/mol/\r{A} and debye,
  respectively. Additionally, the $R^2$ coefficient for the energy is
  shown.}
\label{tab:oos_errors}
\end{table}

\noindent 
The better of the two PES$_{\rm MP2}$ is characterized by
MAE$(E)=0.012$~kcal/mol, RMSE$(E)=0.019$~kcal/mol,
MAE$(F)=0.026$~kcal/mol/\r{A} and
RMSE$(F)=0.195$~kcal/mol/\r{A}. Accuracies similar to PES$_{\rm MP2}$
are achieved for PES$_{\rm TL}$, although the test set (44 geometries)
is smaller than for learning at the MP2 level: the performance is
MAE$(E)=0.007$~kcal/mol, RMSE$(E)=0.016$~kcal/mol,
MAE$(F)=0.076$~kcal/mol/\r{A}, and RMSE$(F)=0.580$~kcal/mol/\r{A}, see
Table~\ref{tab:oos_errors} which also reports results for the dipole
moments.\\

\noindent
The quality of PES$_{\rm MP2}$ can also be assessed in terms of
molecular geometries and the energy barrier for double proton
transfer. The MP2/AVTZ optimized geometries of FAM, FAD and the proton
transfer TS (see Tables~S1-S3)
are reproduced by PES$_{\rm MP2}$ with RMSEs smaller than 0.003~\r{A}.
The energy barrier for PES$_{\rm MP2}$ is 6.69~kcal/mol compared with
6.71~kcal/mol from \textit{ab initio} calculations, i.e. a difference
of 0.02~kcal/mol. The proton transfer barrier, which is underestimated
at the MP2 level of theory, is one of the essential features of the
FAD PES$_{\rm TL}$. Earlier studies at the coupled cluster level of
theory yielded barrier heights of 7.95, 8.16 and 8.30~kcal/mol using
the CCSD(T)-F12a/haDZ, CCSD(T)-F12a/haTZ//CCSD(T)-F12a/haDZ and
CCSD(T)/aV5Z//MP2/aV5Z level of theory,
respectively.\cite{qu2016ab,ivanov2015quantum}. The barrier for DPT at
the CCSD(T)/aug-cc-pVTZ level was found to be
7.9~kcal/mol\cite{MM.fad:2016}. This agrees favourably with a barrier
of 7.92~kcal/mol from the present PES$_{\rm TL}$. The experimental
barrier height is estimated to be 7.3~kcal/mol using a 3D-model to
reproduce the tunneling splitting of 331.2 MHz\cite{caminati:2019}. A
somewhat lower barrier for DPT (7.2 kcal/mol) was also found from
morphing an MP2-based full-dimensional PES to reproduce the
experimentally observed, broad IR absorption from finite-temperature
MD simulations.\cite{MM.fad:2016}\\

\noindent
The dissociation energy $D_e$ on the PES$_{\rm TL}$ for FAD into two
FAMs is --16.79~kcal/mol which is identical to that from explicit
\textit{ab initio} calculations at the CCSD(T)/AVTZ level of
theory\cite{miliordos2015validity}. Because the training set for the
full-dimensional PES contains structures of FAM and FAD, dissociation
along the reaction path connects the minimum energy structures of the
two asymptotic states (2 separate FAMs and FAD, respectively).\\

\subsection{Harmonic Frequencies}
\label{sec:harmonic}
\begin{figure}[h!]
\centering
\includegraphics[width=0.7\textwidth]{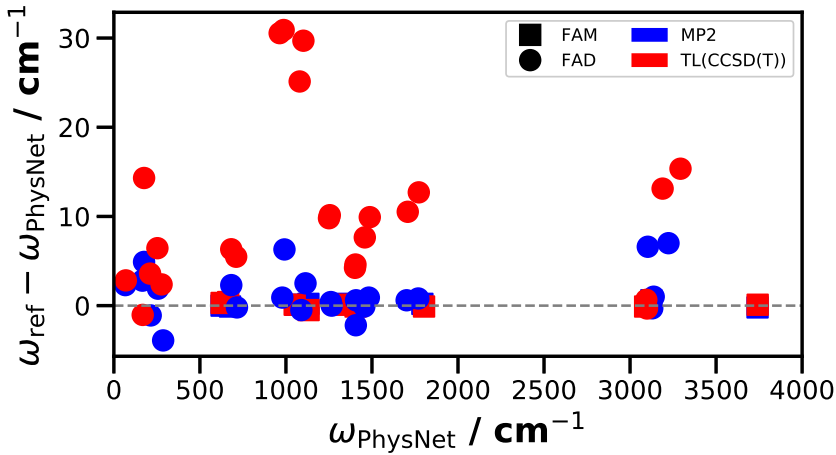}
\caption{The accuracy of harmonic frequencies using PES$_{\rm MP2}$
  and PES$_{\rm TL}$ compared with the corresponding reference
  \textit{ab initio} values. The FAD CCSD(T) values are taken from
  Ref.~\citenum{kalescky2013local}. Almost no error is visible for FAM
  whereas errors up to $\sim 30$~cm$^{-1}$ are found for FAD.}
\label{fig:harmfreq}
\end{figure}

The FAM harmonic frequencies calculated from the two PhysNet PESs are
compared to their respective MP2 and CCSD(T) reference values in
Figure~\ref{fig:harmfreq} (squares) and Table~S4
reports all frequencies. The harmonic frequencies on PES$_{\rm MP2}$
reproduce their reference frequencies with a MAE of 0.2~cm$^{-1}$. It
is noteworthy that PES$_{\rm TL}$ is able to reproduce the
CCSD(T) harmonic frequencies of FAM with a MAE of 0.2~cm$^{-1}$ with
only 425 FAM geometries used in the TL.\\

\noindent
Similarly for FAD, the PhysNet harmonic frequencies are compared with
those from direct {\it ab initio} calculations in
Figure~\ref{fig:harmfreq} (circles) and
Table~S5. Frequencies from PES$_{\rm MP2}$ match the
{\it ab initio} harmonic frequencies with a MAE of 2.1~cm$^{-1}$ and a
maximum unsigned deviation of 7~cm$^{-1}$ for mode 24. For FAD the
harmonic frequencies on PES$_{\rm TL}$ show a somewhat larger MAE than
frequencies from PES$_{\rm MP2}$. The reference CCSD(T)/aug-cc-pVTZ
harmonic frequencies were those from
Ref.~\citenum{kalescky2013local}. It is possible that slightly
different convergence criteria were used in these calculations than
the ones used here. But because such calculations are very time
consuming it was decided not to repeat them here. The MAE between
reference frequencies and those from PES$_{\rm TL}$ is 10.7~cm$^{-1}$
and deviations up to $\sim 30$~cm$^{-1}$ for single frequencies
occur.\\

\subsection{VPT2 Calculations}
VPT2 calculations are carried out for FAM and FAD using PES$_{\rm
  MP2}$ and PES$_{\rm TL}$ as a complement to finite-$T$ spectra, see
Figures~\ref{fig:fam_ir} and \ref{fig:ir_fad} and
Table~\ref{tab:vpt2}. The VPT2 frequencies for FAM obtained from
PES$_{\rm TL}$ agree well with experiment (MAE of 8.4~cm$^{-1}$) for
which the largest deviation is found for the highest frequency mode,
i.e. a difference of 22.2~cm$^{-1}$. The corresponding intensities
capture the general trend of the experiment and, converse to the
finite-$T$ spectra (see below), even the intensities of the two lowest
modes agree with experiment (see Figure~\ref{fig:fam_ir}). For FAD,
the MAE of the VPT2 frequencies is 19.3~cm$^{-1}$ compared with
experiment whereas the intensities agree favourably (see
Figure~\ref{fig:ir_fad}). For FAD the MAEs of the VPT2 calculations on
the two NN-learned PESs with respect to MP2 reference and experiment
are 6.4 cm$^{-1}$ (PES$_{\rm MP2}$) and 19.3 cm$^{-1}$ (PES$_{\rm
  TL}$), respectively. For PES$_{\rm TL}$ the deviations are largest
for the low (modes 1 to 6, MAE of 30.2 cm$^{-1}$) and high frequency
vibrations (modes 22 to 24), see Table \ref{tab:vpt2}.\\

\noindent
One interesting observation concerns the correct
assignment\cite{nejad2021raman} of the fundamental $\nu_5$ (mode 5 in
Table~\ref{tab:vpt2}) and $2\nu_9$ (first overtone of mode 2 in
Table~\ref{tab:vpt2}) modes in FAM. VPT2 calculations using PES$_{\rm
  TL}$ find $\nu_5$ at 1296.0 cm$^{-1}$ and $2\nu_9 =
1211.9$~cm$^{-1}$ which is consistent with recent Raman data measured
in a jet reporting a frequency of 1306 ~cm$^{-1}$ and
1220~cm$^{-1}$ for the fundamental and the overtone,
respectively\cite{nejad2020increasing}. VPT2 calculations at the MP2
level using PES$_{\rm MP2}$, on the other hand, find $\nu_5$ = 1220.6
and $2\nu_9 = 1297.9$~cm$^{-1}$, i.e. the opposite assignment compared
with experiment and calculations at the higher level of theory.\\

\begin{table}[]
\begin{tabular}{c|c|c|c|c||c|c|c|c|}
    & \multicolumn{4}{|c||}{\bf FAM} & \multicolumn{4}{c|}{\bf FAD} \\\hline
 & PES$_{\rm MP2}$ & MP2   &  PES$_{\rm TL}$ & Exp$^{\rm a}$   & PES$_{\rm MP2}$  & MP2   &  PES$_{\rm TL}$ & Exp \\\hline
1	&	619.9	&	619.5	&	621.6	&	626.17	&	78.0	&	70.2	&	91.6	&	69.2	$^{\rm	b}$\\	
2	&	643	&	641.8	&	632.2	&	640.73	&	178.5	&	160.4	&	164.4	&	161.0	$^{\rm	c}$	\\
3	&	1036.7	&	1036.4	&	1029.1	&	1033.47	&	185.0	&	171.0	&	200.9	&	168.5	$^{\rm	b}$\\	
4	&	1097.2	&	1098.2	&	1098.6	&	1104.85	&	206.2	&	197.3	&	229.4	&	194.0	$^{\rm	c}$\\	
5	&	1222.2	&	1220.6	&	1296.0	&	1306.2	&	255.8	&	246.0	&	265.8	&	242.0	$^{\rm	c}$\\	
6	&	1380.7	&	1381.0	&	1374.9	&	1379.05	&	281.1	&	271.9	&	327.6	&	264.0	$^{\rm	d}$\\	
7	&	1761.3	&	1760.5	&	1768.2	&	1776.83	&	686.4	&	678.5	&	679.6	&	682.0	$^{\rm	c}$\\	
8	&	2975.4	&	2967.8	&	2935.3	&	2942.06	&	713.4	&	707.6	&	707.4	&	698.0	$^{\rm	b}$\\	
9	&	3549.8	&	3554.9	&	3548.3	&	3570.5	&	945.2	&	936.0	&	925.6	&	911.0	$^{\rm	e}$	\\
10	&		&	&	&	&				971.6	&	966.1	&	958.1	&	942	$^{\rm	f}$\\	
11	&		&	&	&	&				1065.2$^\ast$	&	1062.6	&	1075.4	&	1050	$^{\rm	g}$\\%	
12	&		&	&	&	&				1083.9$^\ast$	&	1076.6	&	1084.6	&	1060	$^{\rm	h}$\\%	
13	&		&	&	&	&				1237.1	&	1233	&	1238.2	&	1214	$^{\rm	h}$\\%	
14	&		&	&	&	&				1238.4	&	1240.6	&	1240.1	&	1233.9	$^{\rm	i}	$\\%
15	&		&	&	&	&				1370.4	&	1369.2	&	1371.2	&	1371.78	$^{\rm	j}$\\%	
16	&		&	&	&	&				1376.7	&	1373.2	&	1374.4	&	1375	$^{\rm	h}$\\%	
17	&		&	&	&	&				1403$^\ast$	&	1407.6	&	1406.6	&	1415	$^{\rm	g}$\\%	
18	&		&	&	&	&				1426.7$^\ast$	&	1430.7	&	1435	&	1454	$^{\rm	b}$\\	
19	&		&	&	&	&				1662.6	&	1659.1	&	1661.6	&	1666	$^{\rm	k}$\\%	
20	&		&	&	&	&				1732.9	&	1732.9	&	1734.5	&	1741	$^{\rm	k}$\\%	
21	&		&	&	&	&				2763$^\ast$	&	2763.5	&	2875	&	2900	$^{\rm	d}$\\	
22	&		&	&	&	&				2945.4$^\ast$	&	2961.7	&	2912.3	&	2939.7	$^{\rm	l}$\\	
23	&		&	&	&	&				2961.8	&	2962.9	&	2920.1	&	2949	$^{\rm	h}$\\%	
24	&		&	&	&	&				2968.5	&	2963.4	&	3011.2	&	3050	$^{\rm	b}$\\\hline	
{\bf MAE} &   {\bf 2.0}  &       &  {\bf 8.4}  &     &     {\bf 6.4}     &           &  {\bf 19.3}%
\end{tabular}
\caption{VPT2 anharmonic frequencies (in cm$^{-1}$) for FAM and FAD
  calculated using PhysNet trained on MP2 data (PES$_{\rm MP2}$) and
  transfer learned to CCSD(T) (PES$_{\rm TL}$) quality. They are
  compared to their reference \textit{ab initio} values (MP2) and to
  experiment ($^{\rm a}$Ref.~\citenum{nejad2020increasing} and
  Refs. therein; $^{\rm b}$Ref.~\citenum{georges2004jet}; $^{\rm
    c}$Ref.~\citenum{xue2009probing}; $^{\rm
    d}$Ref.~\citenum{suhm:2012}; $^{\rm
    e}$Ref.~\citenum{xue2011raman}; $^{\rm
    f}$Ref.~\citenum{kollipost2015schwingungsdynamik}; $^{\rm
    g}$Ref.~\citenum{millikan1958fad}; $^{\rm
    h}$Ref.~\citenum{bertie:1982}; $^{\rm
    i}$Ref.~\citenum{goroya2014high}; $^{\rm
    j}$Ref.~\citenum{duan:2017}; $^{\rm
    k}$Ref.~\citenum{meyer2018vibrational}; $^{\rm
    l}$Ref.~\citenum{ito2000jet}). The MAEs of PES$_{\rm MP2}$ and
  PES$_{\rm TL}$ are given with respect to the \textit{ab initio} MP2
  and experimental values, respectively. Mode 14 of FAD is involved in
  a strong resonance triad from which a frequency of 1233.9~cm$^{-1}$
  is assigned to the fundamental.  Note that the symmetry of the VPT2
  frequencies marked with a $^\ast$ do not correspond to the
  assignments of the experimental frequencies (see,
  e.g. Ref.~\citenum{georges2004jet} or
  \citenum{kollipost2012communication}). The MAEs of the frequencies
  of PES$_{\rm MP2}$ and MP2 with respect to experiment are 22.0 and
  18.6~cm$^{-1}$, respectively.}
\label{tab:vpt2}
\end{table}

\noindent
For the errors of the VPT2 calculation of FAD there are several
potential reasons. First, PES$_{\rm TL}$ is less accurate in
reproducing harmonic frequencies at the CCSD(T) level. Secondly, the
low frequency harmonic modes 1 to 6 agree well with experiment (see
Table~S5). However, VPT2 determines anharmonic
corrections to be added to or subtracted from the harmonic
frequencies. Hence, if the harmonic frequencies agree well with
experiment, the VPT2 frequencies potentially differ from experimental
values and the disagreement is larger for larger anharmonic
corrections as is often the case for low frequency
vibrations. Thirdly, VPT2 calculations also may incur larger errors,
in particular for proton-bound dimers with large amplitude
motions.\cite{franke2021vpt2} Finally, the above errors can also
accumulate which then leads to worse agreement between experimentally
observed and computed anharmonic frequencies despite the high quality
of PES$_{\rm TL}$. It should be noted that VPT2 calculations at the
CCSD(T) level using conventional electronic structure codes for FAM is
computationally demanding and for FAD it is unfeasible.\\

\subsection{Finite-$T$ Infrared Spectra}
\begin{figure}[h!]
\centering
\includegraphics[width=0.8\textwidth]{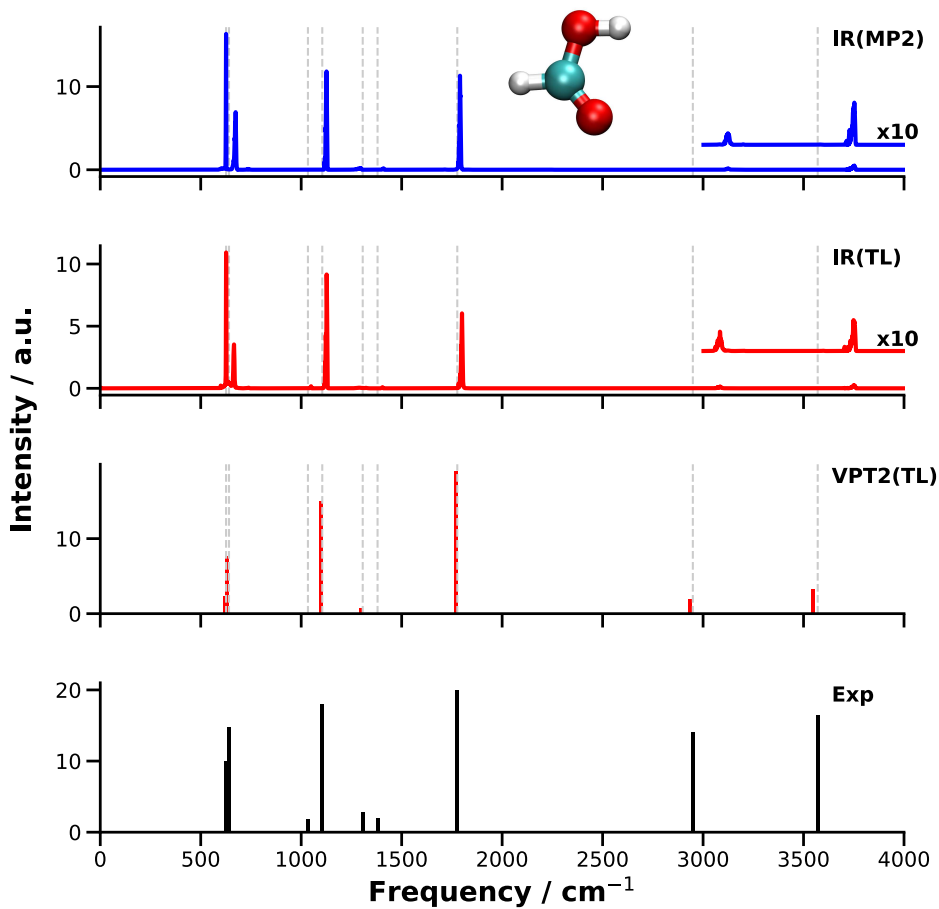}
\caption{Infrared spectrum of FAM obtained from PES$_{\rm MP2}$ (blue)
  and from PES$_{\rm TL}$ (red) using finite-$T$ simulations (IR) and
  the VPT2 approach. The results are complemented with experimental
  frequencies for comparison.\cite{freytes2002overtone,tew2016ab} The
  gray dashed lines indicate the positions of the experimental
  frequencies and the XH peaks are scaled for better readability. The
  intensities of the VPT2 calculation capture the general trend of the
  experimental measurements whereas it is less clear for the
  finite-$T$ spectra. The finite-$T$ spectra (IR) are calculated from
  the dipole-dipole auto-correlation function of 1000 MDs each run for
  200 ps with PES$_{\rm MP2}$ and PES$_{\rm TL}$ at 300~K. }
\label{fig:fam_ir}
\end{figure}

{\bf Formic Acid Monomer:} The IR spectra for FAM obtained from MD
simulations on PES$_{\rm MP2}$ and PES$_{\rm TL}$ are shown in
Figure~\ref{fig:fam_ir} together with the experimental
frequencies\cite{tew2016ab,freytes2002overtone}. Both computed spectra
are averages over 1000 independent $NVE$ MD simulations run at 300~K
for 200~ps each. The low frequency modes faithfully describe those
measured experimentally. On the other hand, the computed high
frequency X-H modes are consistently shifted to the blue by 150 to 200
~cm$^{-1}$ relative to experiment. It is known that specifically for
high-frequency modes the anharmonic regions of the PES are not
sufficiently sampled in finite-$T$ MD
simulations.\cite{qu2018ir,qu2018quantum,MM.oxa:2017} This usually
leads to overestimation of computed frequencies when compared with
experiments as is also found here. The OH stretch frequency for FAM
using PES$_{\rm MP2}$ and PES$_{\rm TL}$ is centered around the same
frequency of $3750$~cm$^{-1}$ which compares with an experimental
value of 3571~cm$^{-1}$.\cite{freytes2002overtone} Counter to the
expectation that on a higher-level PES agreement between computation
and experiment improves, PES$_{\rm MP2}$ and PES$_{\rm TL}$ find the
OH stretch mode at essentially the same frequency. Hence, the inferior
performance of MD for the high-frequency XH stretch modes is not
primarily related to the quality of the PES but rather due to
shortcomings of finite-temperature MD simulations to realistically
sample the anharmonicity of the PES.\\

\begin{figure}[htbp]
\centering
\includegraphics[width=0.9\textwidth]{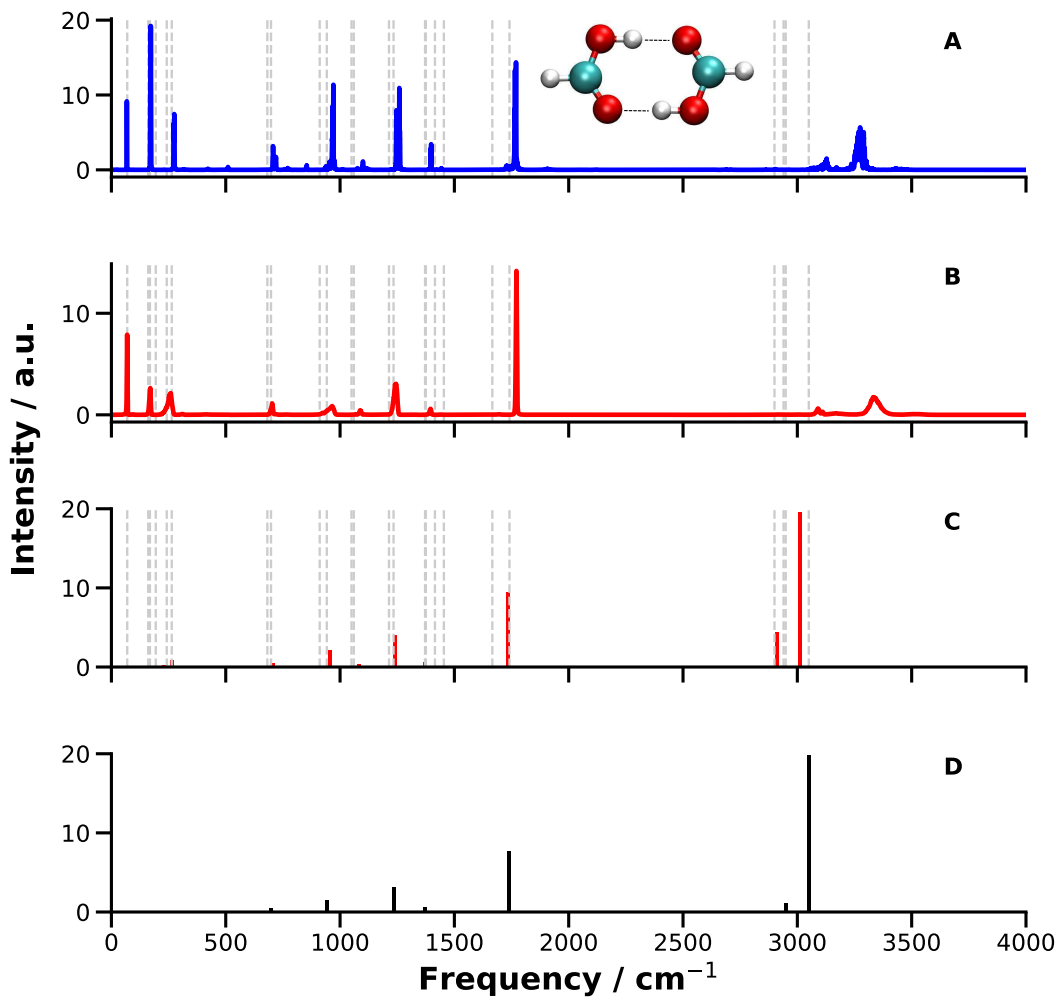}
\caption{Infrared spectrum of FAD obtained from the PES$_{\rm MP2}$
  (panel A) and from the PES$_{\rm TL}$ (panel B) using finite-$T$
  simulations (A, B) and the VPT2 approach (C). The results are
  complemented with experimental frequencies
  \cite{kalescky2013local,suhm:2012} and intensities for 7
  fundamentals\cite{yokoyama1991extended} (D). Six intensities of
  Ref.~\citenum{yokoyama1991extended} were estimated based on
  experimental data from Ref.~\citenum{marechal1987ir} and extended
  with one intensity from Ref.~\citenum{berckmans1988ab}.
  Experimentally, the intensity of the OH-stretch is distributed over
  a much wider frequency range.\cite{georges2004jet} The gray dashed
  lines indicate the positions of the observed fundamental
  frequencies. The finite-$T$ spectra (IR) are calculated from the
  dipole-dipole auto-correlation function of 1000 MDs each run for 200
  ps with the PES$_{\rm MP2}$ and PES$_{\rm TL}$ at 300~K.}
\label{fig:ir_fad}
\end{figure}

\noindent
{\bf Formic Acid Dimer:} The averaged IR spectrum from 1000
independent trajectories, each 200 ps in length using PES$_{\rm MP2}$
and PES$_{\rm TL}$ at 300~K for FAD is shown in
Figure~\ref{fig:ir_fad}. For the modes below 2000~cm$^{-1}$ the
position and the relative intensities from the computations compare
favourably with experiments.\cite{kalescky2013local,suhm:2012} Even
finer details, such as the low intensities of the peaks at 942 and
1233.9~cm$^{-1}$ are surprisingly well captured by simulations with
the two PESs. It is also of interest to note that for some modes the
computed harmonic, anharmonic and experimentally observed frequencies
agree rather well. For example, for the lowest frequency mode of FAD
$\omega({\rm PES_{TL}})=69.5$, $\nu({\rm MD(PES_{TL})})=68.7$, and
$\nu({\rm Exp})=69.2$~cm$^{-1}$\cite{georges2004jet}. Larger
differences between the predicted frequencies and experiment are,
however, again found in the high frequency range. Based on the normal
modes, an OH stretch frequency of 3273 and of 3338~cm$^{-1}$ is found
in the IR of the FAD on PES$_{\rm MP2}$ and PES$_{\rm TL}$,
respectively. The agreement between experiment and simulations is
similar to that found from 12 ps simulations on the PIP-based PES
although there the 1741 cm$^{-1}$ band appears to be shifted somewhat
to the blue.\cite{qu2018ir}\\

\noindent
The experimental Fourier Transform transmittance spectrum of FAD
recorded in a jet reported\cite{georges2004jet} a broad band with
superimposed sharp features extending from below 2600 cm$^{-1}$ up to
at least 3300 cm$^{-1}$. The sharp features arise mostly from
combination bands and make it difficult to assign one specific
absorption feature to the OH stretch vibration. However, it is
reasonable to assume that it is located to the blue side of the
CH-stretch band rather than to the red side of it. Interestingly, the
width of the jet-cooled spectrum is comparable to that recorded at
room temperature. Hence, cooling only leads to sharpening of certain
features but not to a simpler spectrum. This is consistent with other
studies on formic and acetic acid.\cite{zielke:2007,haber:2001} A
potential assignment of the OH stretch vibration in FAD was
made\cite{georges2004jet} to a signature at 3084 cm$^{-1}$ by
comparing with earlier calculations (SCF with double zeta basis
set).\cite{schaefer:1987} Similarly, earlier Raman spectra of
(DCOOH)$_2$ reported a broad absorption around and above 3000
cm$^{-1}$ and assigned features between 2565 cm$^{-1}$ and 3427
cm$^{-1}$ to the dimer OH stretching band.\cite{bertie:1986} Finally,
a broad absorption was also found from lower resolution, room
temperature IR spectroscopy\cite{MM.fad:2016} for which spectral
subtraction techniques were used to obtain the IR spectrum of FAD. The
transition assigned to the OH-stretch was the region between 2600 and
3400 cm$^{-1}$ with a maximum at $\sim 3100$
cm$^{-1}$.\cite{MM.fad:2016} Hence, assignment of the OH-stretch
frequency is not straightforward. Here, a position of $3050 \pm 100$
cm$^{-1}$ was used to compare with. However, it needs to be stressed
that considerable uncertainty about both, the position and the width
of this band exist.\\

\begin{figure}[!h]
\centering
\includegraphics[width=0.9\textwidth]{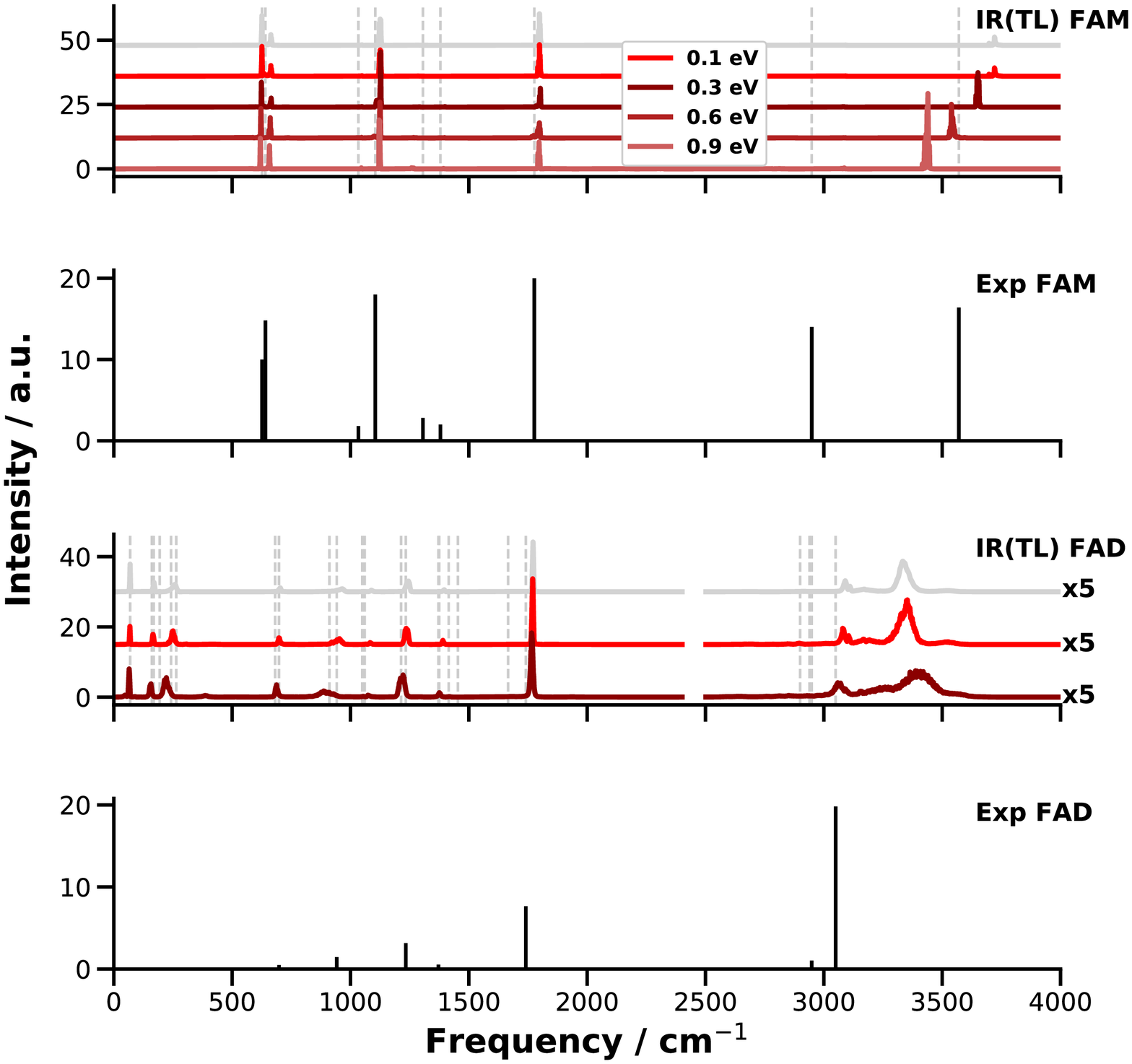}
\caption{IR spectra from $E-$dependent MD simulations for FAM and FAD
  with PES$_{\rm TL}$.  For comparison, for FAM the harmonic ZPE of
  the OH stretch is 0.23~eV and the total ZPEs are $\sim 0.9$ and
  1.9~eV. The grey vertical lines mark the peak positions of the
  experiments\cite{freytes2002overtone,tew2016ab} and the grey spectra
  are calculated from room temperature simulations on PES$_{\rm TL}$
  (see Figs.~\ref{fig:fam_ir} and \ref{fig:ir_fad}). For FAM a clear
  energy dependence is seen: the OH peak $\sim 3725$~cm${^{-1}}$ with
  an energy of 0.1~eV shifts to the red and reaches $\sim
  3435$~cm$^{-1}$ with an energy of 0.9~eV. This energy dependence was
  not achieved for FAD, for which the OH stretch peak was mainly
  broadened and blue shifted for 0.3~eV. In contrast, the C--H peak is
  red shifted as expected. For FAD, the IR signals above
  2500~cm$^{-1}$ are scaled by a factor of 5 for better visibility.}
\label{fig:fam_ir2} 
\end{figure}

\noindent
The failure of finite-$T$ MD simulations to even qualitatively capture
the spectroscopy of high-frequency modes is primarily related to the
fact that at $T\sim300$~K, at which most MD studies are carried out,
the anharmonicities of X--H vibrations are not sampled
adequately. Because constraining the correct amount of ZPE in the
modes of polyatomic molecules is as of now an unsolved
problem\cite{qu2018quantum}, an alternative that has been considered
is to run MD simulations at higher
temperature\cite{qu2018quantum,MM.oxa:2017}. Here, simulations with
increasing amount of energy in the OH stretch of FAM and FAD have been
carried out using PES$_{\rm TL}$. The $NVE$ simulations are run with
excess energies of 0.1, 0.3, 0.6, and 0.9 eV for FAM and of 0.1 and
0.3 eV for FAD in each of the OH stretch modes. For comparison, the
(harmonic) ZPE of FAM in this mode is 1647~cm$^{-1}$ (equivalent to
0.2~eV). This energy is initially placed into the OH stretch(es) as
kinetic energy. The $E$-dependent IR spectra
(Figure~\ref{fig:fam_ir2}) correspond to an ensemble of 45
trajectories run for 200 ps each. For FAM it was found that with
increasing energy content in the OH stretch the associated band shifts
progressively towards the red as expected. With an excess of 0.5 eV
the computed OH stretch aligns with that reported from
experiment.\\

\noindent
Even though the energy redistribution to the lower frequency modes
does occur on the time scale of the simulations the spectroscopic
features below 2000~cm$^{-1}$ do not respond strongly to the increased
temperature. To assess the redistribution of vibrational energy a
supplemental trajectory is run for FAM with an excess of 0.5~eV in the
OH stretch mode. The energy remains in this mode for at least 150 ps
after which it slowly redistributes into other degrees of freedom, see
Figure~S1.\\

\noindent
For FAD running the simulations at higher temperature apparently does
not solve the problem of overestimating the location of the XH
bands. As one additional complication, the binding energy,
i.e. $E_{\rm FAD} - 2E_{\rm FAM}$, of the dimer is only
$-16.79$~kcal/mol ($-0.73$~eV). Hence, excitation of the two OH
stretch vibrations with more than $\sim 8.5$ kcal/mol ($\sim 0.35$ eV)
leads to dissociation of the dimer and the spectroscopy of the
H-bonded OH$\cdots$O motifs can not be probed. For the simulations
with 0.1~eV of excess the OH$\cdots$O hydrogen bond (distance between
O$_{10}$ and H$_5$ or O$_{2}$ and H$_6$, see
Figure~\ref{fig:ooserrors}) extends up to 2.3~\r{A} compared with an
equilibrium separation of 1.68 \AA\/. With an excess of 0.3 eV the two
OH$\cdots$O ``bonds'' elongate up to 3.3~\r{A}. This indicates
considerable destabilization of the dimer due to partial breaking of
the OH bonds and the two FAMs in the dimer start to behave more like
monomers without fully dissociating FAD. Loosening the OH$\cdots$O
contact affects, however, the vibrational frequency of the OH
stretch. Visual inspection also shows that an out of plane bending
motion of the two monomers is excited due to vibrational energy
redistribution which further weakens the OH$\cdots$O hydrogen
bond. For the simulations of FAD with 0.3 eV excess the OH stretch
shifts to the blue by $\sim 70$~cm$^{-1}$ whereas the CH stretch
shifts to the red as expected for simulations at higher temperature.\\

\subsection{Diffusion Monte Carlo and Binding Energy $D_0$}
The ML PESs can also be used to determine the dissociation energy
$D_0$ of FAD from computations. For this, a high-quality estimate for
the total ZPE for FAM and FAD is required which can be obtained from
DMC simulations. The average ZPE from 10 independent DMC simulations
for FAM using the PES$_{\rm TL}$ is $20.932\pm 0.033$~kcal/mol
($7321.1\pm 11.5$~cm$^{-1}$). Similarly for FAD a ZPE of $43.741\pm
0.013$~kcal/mol ($15298.9 \pm 4.5$~cm$^{-1}$) was found. The average
DMC energies and standard deviations are determined from block
averaging\cite{toulouse2016introduction}. These ZPEs (accounting for
anharmonicity) compare also favourably with 20.900 and
43.885~kcal/mol, respectively, for the monomer and the dimer from the
VPT2 calculations using the same PES (PES$_{\rm TL}$). Similarly, the
result for FAD also agrees well with a recent study using a PES based
on permutationally invariant polynomials at the CCSD(T)-F12a/haTZ
level of theory and DMC calculations which reported a ZPE of $15337
\pm 7$ cm$^{-1}$ ($43.850 \pm 0.020$~kcal/mol).\cite{qu2016ab}\\

\noindent
Based on these ZPEs the FAD dissociation energy $D_0 = D_e^C -2{\rm
  ZPE}_{\rm FAM} + {\rm ZPE}_{\rm FAD}$ can be determined. Here,
$D_e^C$ is the corrected (see below) binding energy (``from the bottom
of the potential'') and ${\rm ZPE}_{\rm FAM}$ and ${\rm ZPE}_{\rm
  FAD}$ are the ZPEs of FAM and FAD, respectively. The dissociation
energy $D_e$ needs to be corrected a) to account for basis set
superposition error (BSSE) and b) to include effects due to
finite-size basis set effects by extrapolation to the complete basis
set (CBS) limit. These corrections were determined in previous
work\cite{miliordos2015validity} and will be used here. This is
possible because $D_e = -16.79$ kcal/mol from the present work for
dissociation of FAD into two FAMs each at their respective equilibrium
structure, is identical to the value reported from the earlier
CCSD(T)/aug-cc-pVTZ calculations.\cite{miliordos2015validity}
Including corrections for BSSE at the CBS limit yields $D_e^C =
-16.11$~kcal/mol.\cite{miliordos2015validity} Together with the ZPEs
of FAM and FAD determined above this yields a best estimate of $D_0 =
-14.23 \pm 0.08$ kcal/mol where the error is that of the ZPEs from the
DMC calculations. This value is consistent, within error bars, with
the experimentally reported value of $-14.22 \pm
0.12$~kcal/mol\cite{suhm:2012} and compares with $D_0 = -14.3 \pm 0.1$
kcal/mol from computations with ZPEs obtained from a hybrid VPT2
approach (harmonic frequencies from CCSD(T)/AVQZ and correcting for
anharmonic effects using MP2/AVDZ).\cite{miliordos2015validity}\\

\section{Discussion and Conclusion}
In this work the harmonic and anharmonic vibrational spectra for FAM
and FAD in the gas phase have been determined based on correctly
dissociating machine-learned PESs at the MP2 level of theory and
transfer-learned to the CCSD(T) level of theory with the aug-cc-pVTZ
basis set. Harmonic frequencies on PES$_{\rm MP2}$ and PES$_{\rm TL}$
from diagonalizing the Hessian are in good agreement with those
determined from conventional {\it ab initio} calculations. Accounting
for anharmonicity in the vibrations through VPT2 calculations yields
spectra in good agreement with those observed experimentally. The
finite-temperature MD simulations find good agreement with
experimental spectra for all modes below 2000 cm$^{-1}$ but disagree
by several 100 cm$^{-1}$ for the high-frequency XH stretch
vibrations. This confirms earlier findings and is related to the fact
that MD simulations at 300 K do not sample the mechanical
anharmonicity of XH stretch or any other strongly anharmonic modes
sufficiently to give reliable frequencies from such an
approach. Contrary to that, a 24-mode VSCF/VCI calculation on a PES
calculated at the CCSD(T)-F12a/haTZ level of theory and fit to
permutationally invariant polynomials\cite{qu2016ab} correctly
describes the spectroscopy and increased width of the OH-stretch
band.\cite{qu2018quantum} \\

\begin{table}[h]
\begin{tabular}{l || l | l || l | l | l || l}
  [cm$^{-1}$] &$\omega$(MP2) & $\omega$(CCSD(T)) & $\nu$(MP2) & $\nu$(TL) & VPT2(TL) & Exp \\
  \hline
  OH (FAM) & 3741 & 3742 & 3750 & 3750 & 3548 & 3571$^{\rm a}$ \\
  OH (FAD) & 3230 & 3309\cite{kalescky2013local} / 3306 \cite{miliordos2015validity} & 3273 & 3338 & 3011 & $3050 \pm 100^{\rm b}$ \\
  \hline
  $\Delta$ & --511 & --433 & --477 & --412 & --537 & --$521 \pm 100$
\end{tabular}
\caption{OH stretch frequencies of FAM and FAD obtained from
  \textit{ab initio} harmonic frequency calculations ($\omega$(MP2)
  and $\omega$(CCSD(T))), from MD simulations at 300 K on PES$_{\rm
    MP2}$ ($\nu$(MP2)) and PES$_{\rm TL}$ ($\nu$(TL)), and from VPT2
  calculations using PES$_{\rm TL}$, compared with experiment ($^{\rm
    a}$Ref.~\citenum{nejad2020increasing}; $^{\rm
    b}$Ref.~\citenum{georges2004jet}). CCSD(T) frequencies for the
  dimer are from References \citenum{miliordos2015validity} and
  \citenum{kalescky2013local} and were calculated using the AVTZ and
  AVQZ basis sets, respectively. The position of the H-transfer band
  ($3050 \pm 100^{\rm b}$) is an estimate based on
  Ref.\citenum{georges2004jet}, see also discussion in the text.}
\label{tab:redshift}
\end{table}

\noindent
The present work also allows to make contact with a recent
assessment\cite{suhm:2020} of various sources of errors in the
spectroscopy and thermodynamics of FAD which reported that ``More
often than not, the overestimation of harmonic downshifts in DFT is
qualitatively compensated by the inability of classical dynamics to
sample the anharmonic region probed by the quantum nature of the
hydrogen atom. This frequently provides right answers for the wrong
reasons whenever high-frequency XH stretching spectra are simulated,
because in this case, temperatures of several 1000 K would be needed
to sample the relevant fundamental vibrational displacements.'' 
While some of these statements are supported by the present
results, others require amendments.\\

\noindent
In agreement with this assessment the present work finds that energies
up to 0.5 eV ($\sim 5800$~K) are required for the OH stretch vibration
in FAM to agree with experiment. This is consistent with the
temperature $T = \frac{hc\omega_e}{k}$ required to reproduce the
stretching fundamental of a simple Morse oscillator with harmonic
wavenumber $\omega_e$.\cite{suhm2013femtisecond} On the other hand,
the CCSD(T)-level quality of PES$_{\rm TL}$ allows to probe whether or
not accidental agreement with experiment is found for the complexation
induced red shifts $\Delta \omega$ and $\Delta \nu$. For the harmonic
frequencies, the red shifts are $\Delta \omega = -511$ cm$^{-1}$ and
$\Delta \omega = -433$ cm$^{-1}$ from normal mode calculations at the
MP2 and CCSD(T) levels of theory, compared with $-518$ and $-449$
cm$^{-1}$ from normal modes on the respective NN-learned PES$_{\rm
  MP2}$ and PES$_{\rm TL}$, see Tables~S4 and
S5. MD simulations with PES$_{\rm MP2}$ at 300 K
find $\Delta \nu = -477$~cm$^{-1}$ compared with $-521 \pm
100$~cm$^{-1}$ estimated from experiments\cite{georges2004jet}, see
Table~\ref{tab:redshift}. With PES$_{\rm TL}$ the red shift is $\Delta
\nu = -412$ cm$^{-1}$ and the VPT2 calculations yield a shift of --537
cm$^{-1}$. It should be noted that VPT2 calculations are known to have
limitations for proton bound dimers.\cite{franke2021vpt2} This is also
consistent with the present finding that the ZPEs of FAD from DMC
simulations and VPT2 calculations using PES$_{\rm TL}$ differ by 0.14
kcal/mol (50 cm$^{-1}$) whereas that for FAM is virtually identical.\\

\noindent
The shifts from finite-$T$ MD simulations still suffer from the
limitation that the anharmonicity of the OH stretch is not
sufficiently sampled, irrespective of the level of theory at which the
PES was determined. However, it is likely that the magnitude and the
direction of the error in the anharmonic OH stretch frequencies differ
for FAM compared with FAD. This can be seen most clearly in Figure
\ref{fig:fam_ir2} where the OH stretch in FAM/FAD shifts to the
red/blue with increasing internal energy, respectively. The
complexation induced red shift from simulations on PES$_{\rm MP2}$ is
within 44 cm$^{-1}$ of the experimentally observed value whereas that
on (higher-level) PES$_{\rm TL}$ is lower by 109 cm$^{-1}$. Given the
close agreement for the anharmonic OH-stretch for FAM from simulations
on the two PESs it is expected that PES$_{\rm TL}$ for FAD is of
similar quality as for FAM and only minor improvement is expected by
adding additional reference points in the transfer learning. This is
supported by the finding that for the OH stretch $\Delta \omega$ from
MP2 calculations differs from that on PES$_{\rm MP2}$ by 7 cm$^{-1}$
whereas for CCSD(T) and PES$_{\rm TL}$ the difference is 16
cm$^{-1}$. Likewise, using a larger basis set, such as aug-cc-pVQZ, is
unlikely to change $\omega$ and $\nu$ of the OH stretch appreciably
because the harmonic frequencies from CCSD(T)/aug-cc-pVTZ and
CCSD(T)/aug-cc-pVQZ differ only by 3 cm$^{-1}$, see Table
\ref{tab:redshift}. Thus, the apparent ``agreement'' between
experiment and spectra computed from the present MD simulations on
PES$_{\rm MP2}$ is due to differential undersampling the anharmonicity
of the OH coordinate at 300 K for FAM and FAD and also due to the
uncertainty in assigning the OH-stretch in FAD more definitively.\\

\noindent
This changes as soon as DPT occurs, because then the full
anharmonicity of the energy function along the XH coordinate must be
sampled. From simulations with multiple forward and backward crossings
in the MD simulations it is expected that the anharmonicity is
comprehensively sampled which should provide more realistic IR
spectra. Previous work on FAD accomplished this by using the MMPT PES
for which trajectories 250 ns in length were carried out which sampled
25 DPT events.\cite{MM.fad:2016} Such long simulation times become
accessible because these energy functions can be evaluated at the
speed of a conventional empirical force field. Furthermore, the MMPT
function used did not allow FAD to dissociate into two monomers.\\

\noindent
In summary, the present work uses machine-learned, full dimensional
and reactive PESs at the MP2 and CCSD(T) levels of theory for FAM and
FAD to characterize their vibrational dynamics. It is established that
for framework vibrational modes (below 2000 cm$^{-1}$) computed
frequencies from VPT2 and finite-$T$ MD simulations agree well with
experiments. However, for the high-frequency XH stretch modes - in
particular the OH modes - MD simulations do not sufficiently sample
the anharmonicities which leads to a considerable overestimation of
the experimentally observed frequencies. Hence, it is primarily the
use of MD simulations that is the source of errors for the
disagreement between experiments and simulations. The estimated
experimentally observed red shift of $-521 \pm 100$~cm$^{-1}$ for the
OH stretch peak\cite{georges2004jet} compares with --477~cm$^{-1}$
from finite-$T$ simulations on PES$_{\rm MP2}$ and --412 cm$^{-1}$
when using PES$_{\rm TL}$ in the simulations. The red shift from the
VPT2 calculations is --537 cm$^{-1}$. However, there is considerable
uncertainty on the position and width of the OH-stretch fundamental
which makes it less suited for direct comparison with computations.\\

\noindent
For FAM, running simulations at elevated temperatures does not affect
the framework modes but shifts the OH stretch modes towards the
frequencies observed in experiments. On the other hand, for FAD
destabilization of the dimer prevents using such an approach. Hence,
computing accurate complexation-induced spectral shifts remain a
challenge even on high-level PESs due to shortcomings in the dynamics
simulations underlying the spectroscopy. This is also not expected to
change if approximate quantum methods such as QCMD or RPMD simulations
are used, as was shown explicitly.\cite{MM.oxa:2017,qu2018ir} The
computed dissociation energy of $D_0 = -14.23 \pm 0.08$ kcal/mol from
a combination of electronic structure calculations and DMC simulations
on the CCSD(T)-quality PES$_{\rm TL}$ is in favourable agreement with
the value of $-14.22 \pm 0.1$ kcal/mol from experiments.\\

\section{Acknowledgment}
We thank Prof. J. M. Bowman and Dr. C. Qu for fruitful exchange on the
implementation of DMC and Prof. M. Suhm and Dr. A. Nejad for valuable
correspondence. This work was supported by the Swiss National Science
Foundation through grants 200021-117810, 200020-188724 and the NCCR
MUST, and the University of Basel.

\bibliography{references}
\end{document}

\section{Molecular Geometries}
\begin{table}[h!]
\centering
\begin{tabular}{l|l|l|l}
FAM      &   MP2/AVTZ                   &   ENERGY=-189.48676129            &    \\\hline
 & X &        Y       &        Z       \\\hline
C        & 1.4520860701         & 0.2420155948  & -0.0706270260 \\
H        & 0.3886055057         & -0.0022636281 & -0.0180112547 \\
O        & 1.7795679729         & 0.5699513997  & -1.3349111851 \\
O        & 2.2170078996         & 0.2215156425  & 0.8605086772  \\
H        & 2.7281099589         & 0.7770183078  & -1.3284469693
\end{tabular}\caption{Coordinates and energy (in hartree) of FAM optimized
at the MP2/AVTZ level of theory.}\label{sitab:fam_geom}
\end{table}

\begin{table}[h!]
\centering
\begin{tabular}{l|l|l|l}
FAD           &   MP2/AVTZ                   &  ENERGY=-379.00025498   &         \\\hline
 & X &        Y       &        Z       \\\hline
C               & -1.9764324596        & -0.1494375270 & 0.1666266080  \\
H               & -3.0549779301        & -0.2186023083 & 0.3193894061  \\
O               & -1.6825757966        & 0.2969079440  & -1.0328347642 \\
O               & -1.1776513399        & -0.4610563717 & 1.0396060847  \\
H               & -0.6892121927        & 0.3479665039  & -1.1351578367 \\
C               & 1.7585514984         & 0.1213186293  & -0.4302037094 \\
H               & 2.8370616671         & 0.1933717275  & -0.5818836749 \\
O               & 1.4647336363         & -0.3267641096 & 0.7686141982  \\
O               & 0.9597539040         & 0.4336624956  & -1.3028922022 \\
H               & 0.4714710032         & -0.3780585438 & 0.8709677203 
\end{tabular}\caption{Coordinates and energy (in hartree) of FAD optimized
at the MP2/AVTZ level of theory.}\label{sitab:fad_geom}
\end{table}

\begin{table}[h!]
\centering
\begin{tabular}{l|l|l|l}
FAD TS              & MP2/AVTZ & ENERGY=-378.98955690 &               \\\hline
 & X &     Y          &     Z          \\\hline
C               & 1.7580863319         & 0.0022963293  & 0.0961466133  \\
H               & 2.8483449275         & 0.0047147133  & 0.1321411491  \\
O               & 1.1960765375         & -1.1283848924 & 0.0522983349  \\
O               & 1.1894021598         & 1.1304667771  & 0.1026492092  \\
H               & -0.0079795935        & -1.0950129011 & 0.0132873913  \\
O               & -1.2118222166        & -1.1337245552 & -0.0273401908 \\
O               & -1.2185020156        & 1.1251240211  & 0.0231035418  \\
C               & -1.7805078142        & -0.0055548519 & -0.0208134067 \\
H               & -0.0144445979        & 1.0917544630  & 0.0620743374  \\
H               & -2.8707647037        & -0.0079727954 & -0.0568725511
\end{tabular}\caption{Coordinates and energy (in hartree) of the TS for double proton
transfer optimized
at the MP2/AVTZ level of theory.}\label{sitab:fadts_geom}
\end{table}
\clearpage
\section{Harmonic Frequencies}

\begin{table}[H]
\begin{tabular}{l|l|l||l|l||l}
\textbf{Mode}& \textbf{PES$_{\rm MP2}$}  & \textbf{MP2}& \textbf{PES$_{\rm TL}$}  &\textbf{CCSD(T)}& \textbf{Exp}~\cite{tew2016ab}\\\hline
\textbf{1}	&	625.9	&	625.9	&	626.2	&	626.5	&	626.2	\\
\textbf{2}	&	675.2	&	675.1	&	664.5	&	664.9	&	640.7	\\
\textbf{3}	&	1058.6	&	1058.7	&	1050.9	&	1051.0	&	1033.5	\\
\textbf{4}	&	1130.6	&	1130.8	&	1131.8	&	1131.3	&	1104.9	\\
\textbf{5}	&	1301.5	&	1301.7	&	1310.7	&	1310.8	&	1306.2	\\
\textbf{6}	&	1409.0	&	1409.0	&	1404.8	&	1404.7	&	1380.0	\\
\textbf{7}	&	1793.0	&	1793.2	&	1802.7	&	1802.6	&	1776.8	\\
\textbf{8}	&	3123.6	&	3124.2	&	3087.7	&	3087.6	&	2942.0	\\
\textbf{9}	&	3741.0	&	3740.8	&	3741.7	&	3741.8	&	3570.5	\\\hline

\textbf{MAE}	&	\textbf{0.2}	&	&	\textbf{0.2}						
               
\caption{Comparison of the normal mode frequencies of FAM calculated from PhysNet
  trained on MP2 data (PES$_{\rm MP2}$) and transfer learned to CCSD(T) quality (PES$_{\rm TL}$), the corresponding reference values (MP2 and CCSD(T)) and experimental fundamental
  vibration frequencies from Ref.~\citenum{nejad2020increasing} and Refs. therein. PhysNet reproduces its reference frequencies with a MAE of 0.2~cm$^{-1}$. Largest deviations to
  experiments are found for the O--H and C--H modes having high
  frequencies. All frequencies are given
  in cm$^{-1}$.}\label{sitab:harm_fam}
\end{tabular}
\end{table}

\begin{table}[]
\begin{tabular}{l|l|l||l|l||l}
\textbf{Mode} & \textbf{PES$_{\rm MP2}$} & \textbf{MP2} & \textbf{PES$_{\rm TL}$} & \textbf{CCSD(T)\cite{kalescky2013local}} 
& \textbf{Exp} \\\hline
\textbf{1}	&	66.5	&	68.8	&	69.5	&	72.3	&	69.2	$^{\rm	b}$\\	
\textbf{2}	&	166.0	&	168.8	&	167.5	&	166.5	&	161.0	$^{\rm	c}$	\\
\textbf{3}	&	175.4	&	180.3	&	175.6	&	189.9	&	168.5	$^{\rm	b}$\\	
\textbf{4}	&	214.0	&	212.9	&	210.5	&	214.1	&	194.0	$^{\rm	c}$\\	
\textbf{5}	&	256.5	&	258.4	&	253.0	&	259.4	&	242.0	$^{\rm	c}$\\	
\textbf{6}	&	286.4	&	282.5	&	277.2	&	279.6	&	264.0	$^{\rm	d}$\\	
\textbf{7}	&	682.3	&	684.6	&	682.1	&	688.4	&	682.0	$^{\rm	c}$\\	
\textbf{8}	&	715.3	&	715.1	&	710.7	&	716.2	&	698.0	$^{\rm	b}$\\	
\textbf{9}	&	978.9	&	979.8	&	963.5	&	994.1	&	911.0	$^{\rm	e}$	\\
\textbf{10}	&	991.4	&	997.7	&	986.7	&	1017.6	&	942	$^{\rm	f}$\\	
\textbf{11}	&	1089.6	&	1089.1	&	1079.9	&	1105.0	&	1050	$^{\rm	g}$\\%	
\textbf{12}	&	1113.8	&	1116.3	&	1101.5	&	1131.2	&	1060	$^{\rm	h}$\\%	
\textbf{13}	&	1261.0	&	1261.4	&	1250.2	&	1260.0	&	1214	$^{\rm	h}$\\%	
\textbf{14}	&	1265.9	&	1265.9	&	1254.9	&	1265.0	&	1233.9	$^{\rm	i}	$\\%
\textbf{15}	&	1406.2	&	1404.0	&	1401.5	&	1405.7	&	1371.78	$^{\rm	j}$\\%	
\textbf{16}	&	1407.5	&	1408.1	&	1404.1	&	1408.7	&	1375	$^{\rm	h}$\\%	
\textbf{17}	&	1455.6	&	1455.5	&	1458.3	&	1466.0	&	1415	$^{\rm	g}$\\%	
\textbf{18}	&	1481.8	&	1482.7	&	1487.3	&	1497.2	&	1454	$^{\rm	b}$\\	
\textbf{19}	&	1701.6	&	1702.2	&	1708.0	&	1718.5	&	1666	$^{\rm	k}$\\%	
\textbf{20}	&	1769.1	&	1769.9	&	1772.2	&	1784.9	&	1741	$^{\rm	k}$\\%	
\textbf{21}	&	3103.1	&	3109.7	&	3094.4	&	3095.1	&	2900	$^{\rm	d}$\\	
\textbf{22}	&	3128.9	&	3128.6	&	3098.4	&	3098.1	&	2939.7	$^{\rm	l}$\\	
\textbf{23}	&	3137.4	&	3138.4	&	3189.2	&	3202.3	&	2949	$^{\rm	h}$\\%	
\textbf{24}	&	3223.3	&	3230.3	&	3293.1	&	3308.5	&	3050	$^{\rm	b}$\\\hline	

\textbf{MAE}                           & {\bf 2.1}  &  &  {\bf 10.7}   &  &           
\caption{Comparison of the normal mode frequencies of dimeric FA
  calculated from PhysNet trained on MP2 data (PES$_{\rm MP2}$) and
  transfer learned to CCSD(T) quality (PES$_{\rm TL}$), the
  corresponding reference values (MP2 and CCSD(T), the latter taken
  from Ref.~\citenum{kalescky2013local}) and experimental fundamental
  vibration frequencies ($^{\rm b}$Ref.~\citenum{georges2004jet};
  $^{\rm c}$Ref.~\citenum{xue2009probing}; $^{\rm
    d}$Ref.~\citenum{suhm:2012}; $^{\rm
    e}$Ref.~\citenum{xue2011raman}; $^{\rm
    f}$Ref.~\citenum{kollipost2015schwingungsdynamik} $^{\rm
    g}$Ref.~\citenum{millikan1958fad}; $^{\rm
    h}$Ref.~\citenum{bertie:1982} $^{\rm
    i}$Ref.~\citenum{goroya2014high}; $^{\rm
    j}$Ref.~\citenum{duan:2017}; $^{\rm
    k}$Ref.~\citenum{meyer2018vibrational}; $^{\rm
    l}$Ref.~\citenum{ito2000jet}). The PhysNet frequencies reproduce
  their MP2 reference values with and MAE of 2.1~cm$^{-1}$ and the
  CCSD(T) with a MAE of 10.7~cm$^{-1}$. All frequencies are given in
  cm$^{-1}$.}
\label{sitab:fad_harm}
\end{tabular}
\end{table}
\clearpage
\begin{figure}[htbp]
\centering
\includegraphics[width=0.9\textwidth]{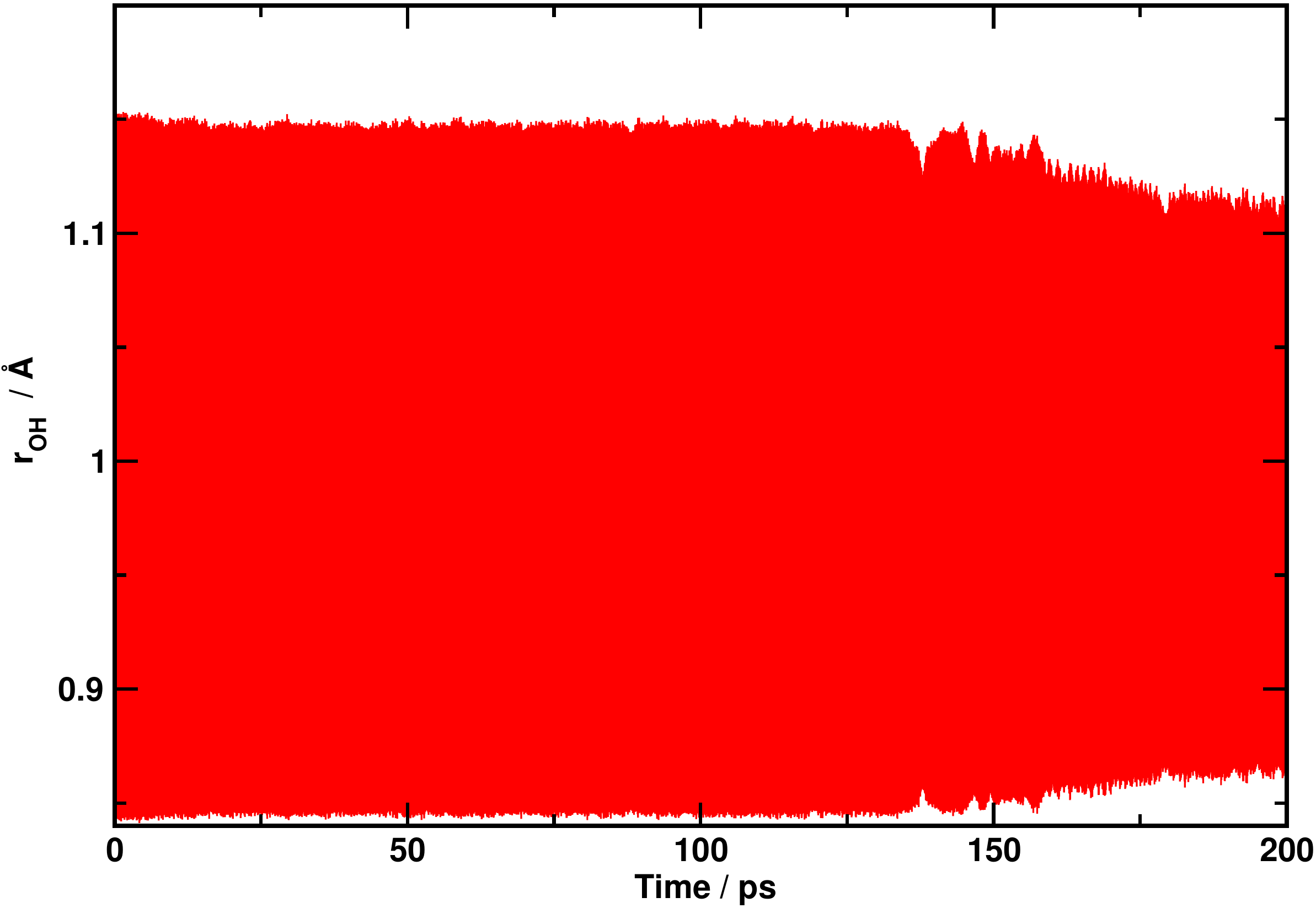}
\caption{MD simulation, 200~ps in length, of FAM run with PES$_{\rm TL}$. The simulation is initialized with a kinetic energy of 0.5~eV in the O--H stretch mode and is used to asses the amount of
vibrational energy redistribution.}
\label{sifig:ivr}
\end{figure}
\clearpage
\bibliography{references}